\documentclass[journal]{IEEEtran}
\usepackage{cite}

\ifCLASSINFOpdf 
\usepackage[pdftex]{graphicx}
\else
\usepackage[dvips]{graphicx}
\fi

\usepackage[cmex10]{amsmath}
\usepackage{algorithmic}
\usepackage[ruled,norelsize]{algorithm2e}

\usepackage{array}

\usepackage{mdwmath}
\usepackage{mdwtab}
\usepackage{todonotes}

\usepackage{eqparbox}

\ifCLASSOPTIONcompsoc
\usepackage[caption=false,font=normalsize,labelfont=sf,textfont=sf]{subfig}
\else
\usepackage[caption=false,font=footnotesize]{subfig}
\fi

\usepackage{fixltx2e}
\usepackage{url}
\usepackage{amsfonts}
\usepackage{amssymb}
\usepackage{amsthm}
\usepackage{bm}
\usepackage[utf8]{inputenc}
\usepackage{booktabs}
\usepackage{url}
\usepackage{cite}
\usepackage{flushend}
\usepackage{epstopdf}

\newcounter{tempEquationCounter} 
\newcounter{thisEquationNumber}

\begin{document}
%

\title{High-Performance Industrial Wireless: Achieving Reliable and Deterministic Connectivity over \\IEEE 802.11 WLANs }




\author{
 Adnan~Aijaz
\thanks{The author is with the Bristol Research and Innovation Laboratory, Toshiba Research Europe Ltd., Bristol, BS1 4ND, U.K. Contact e-mail: adnan.aijaz@toshiba-trel.com}}

\maketitle
\begin{abstract}
\boldmath
Communication for control-centric industrial applications is characterized by the requirements of very high reliability, very low and deterministic latency and high scalability. Typically, IEEE 802.11-based wireless local area networks (WLANs), also known as Wi-Fi networks, are deemed ineligible for industrial control applications owing to insufficient reliability and non-deterministic latency. This paper proposes a novel solution for providing reliable and deterministic communication, through Wi-Fi, in industrial environments. The proposed solution, termed as \textsf{HAR\(^\text{2}\)D-Fi} (\underline{H}ybrid channel \underline{A}ccess with \underline{R}edundancy for \underline{R}eliable and \underline{D}eterministic Wi-\underline{Fi}), adopts hybrid channel access mechanisms for achieving deterministic communication. It also provides temporal redundancy for enhanced reliability. \textsf{HAR\(^\text{2}\)D-Fi} implements different medium access control (MAC) designs that build on the standard physical (PHY) layer. Such designs can be classified into two categories: (a) MAC designs with pre-defined (physical) time-slotted schedule, and (b) MAC designs with virtual time-slotted schedule. Performance evaluation, based on analysis and system-level simulations, demonstrates the viability of \textsf{HAR\(^\text{2}\)D-Fi} for control-centric industrial applications.
\end{abstract}


\begin{IEEEkeywords}
IEEE 802.11, industrial network, MAC, OFDMA, wireless control, WLANs. 
\end{IEEEkeywords}

%
\IEEEpeerreviewmaketitle

\section{Introduction}
\IEEEPARstart{C}{ommunication} in \textit{industrial control networks} \cite{emer_ind_net}  demands very high reliability, very low latency with deterministic guarantees and high scalability for supporting a large number of devices (sensors, actuators, etc.) \cite{ind_cont_net}. The emerging Industry 4.0 \cite{ind4} 
framework envisions several new control-centric applications that connect people, objects and processes in real-time \cite{TI_PIEEE}. 
Existing industrial control networks are  based on wired technologies (fieldbus systems \cite{fieldbus}, industrial Ethernet \cite{ind_ethernet}, etc.). The use of wireless technologies for control applications is still in infancy. Current wireless technologies like ZigBee, WirelessHART, and ISA 100.11a, which are based on  IEEE 802.15 standards, are mainly employed for monitoring applications. \textcolor{black}{Some recent wireless solutions \cite{enclose, WISA, gallop} have been designed for control-centric industrial applications; however these solutions are based on low-power wireless technologies like Bluetooth which yields limited coverage and data rates. } High-performance wireless technologies are not only beneficial for industrial communication but also crucial in the envisioned transformation towards Industry 4.0 \cite{TI_PIEEE}, \cite{f_ind_comm}.

IEEE 802.11-based wireless local area networks (WLANs), also known as Wi-Fi networks,  offer various benefits for industrial communication.  Wi-Fi provides better coverage and supports higher data rates as compared to IEEE 802.15-based solutions. The physical (PHY) layer of legacy Wi-Fi is based on orthogonal frequency division multiplexing (OFDM) which provides robustness against multi-path fading. \textcolor{black}{The PHY layer in the next generation Wi-Fi standard, i.e., IEEE 802.11ax \cite{11ax} is based on orthogonal frequency division multiple access (OFDMA) which provides additional benefits for industrial communication. First, it makes transmissions more resilient to frequency selective fading and interference. Second, it allows partitioning of wide channels into smaller channels which is ideal for low-bandwidth industrial applications. Third, it enables parallel transmissions from multiple users which reduces latency. Last, but not least, it allows more efficient use of resources as compared to OFDM.  } 
Typically, Wi-Fi is deemed ineligible for control-centric industrial communication due to lack of determinism and insufficient reliability. These limitations arise due to randomness and contention-based nature of the medium access control (MAC) layer, which is based on IEEE 802.11 \textit{distributed coordination function} (DCF) i.e., carrier sense multiple access with collision avoidance (CSMA/CA).

The determinism of Wi-Fi can be improved through  a scheduling-like MAC layer functionality. Therefore, techniques like IEEE 802.11 \textit{point coordination function} (PCF) and time division multiple access (TDMA) become particularly attractive. The reliability of Wi-Fi can be improved through packet-level redundancy techniques, for example using the  \textit{parallel redundancy protocol} (PRP), defined in IEC 62439-3 specification \cite{IEC_PRP}. 
The fundamental principle of PRP is to send two copies of the same packet/frame on two distinct networks. The first copy that arrives at the destination is retained, whereas the second  is discarded. PRP provides seamless  redundancy that improves reliability. \textcolor{black}{For adoption in Wi-Fi, PRP requires devices with dual radios, which may not always be feasible for low-cost sensors and actuators in industrial environments.} Besides, it requires a packet/frame duplication entity at higher layers of the protocol stack for packet duplication and duplicate detection.



This paper proposes a high-performance IEEE 802.11-based industrial wireless solution for reliable and deterministic communication. The proposed solution, termed as \textsf{HAR\(^\text{2}\)D-Fi}\footnote{\underline{H}ybrid channel \underline{A}ccess with \underline{R}edundancy for \underline{R}eliable and \underline{D}eterministic Wi-\underline{Fi}}, adopts  hybrid channel access schemes for deterministic communication. It also implements a redundancy mechanism for enhanced reliability. More importantly, \textsf{HAR\(^\text{2}\)D-Fi} provides reliable and deterministic communication \textcolor{black}{\textit{without relying on dual radios}}. \textsf{HAR\(^\text{2}\)D-Fi} has been designed for control-centric industrial applications, for example, field-level communication in factory automation or wireless control of mobile robots in warehouses. The design requirements include support for up to \(100\) devices while providing high reliability and low latency. The key contributions are  summarized as follows.
\begin{itemize}
\item \textcolor{black}{We design \textsf{HAR\(^\text{2}\)D-Fi} for industrial control applications. \textsf{HAR\(^\text{2}\)D-Fi} design (Sections \ref{sect_ka}) incorporates various distinguishing features that are crucial in achieving reliable and deterministic communication. \textsf{HAR\(^\text{2}\)D-Fi} implements different MAC designs which can be classified into two categories: (a) MAC designs with pre-defined (physical) time-slotted schedule over OFDM-based PHY layer, and (b) MAC designs with virtual time-slotted schedule based on the capabilities of OFDMA-based PHY layer.}

\item \textcolor{black}{We develop an analytic framework (Section \ref{sect_af}) for \textsf{HAR\(^\text{2}\)D-Fi} by integrating spatial and protocol aspects and derive closed-form expressions for different performance metrics. }

\item \textcolor{black}{We conduct a comprehensive performance evaluation (Section \ref{sect_pe}) of \textsf{HAR\(^\text{2}\)D-Fi} based on numerical and simulation studies. 
}
\end{itemize}


\section{Related Work} \label{sect_rw}
Improving the reliability and determinism of Wi-Fi networks has been the focus of some recent studies. Cena \textit{et al.} \cite{Wi-Red} proposed Wi-Red which extends the concept of PRP to Wi-Fi networks. Wi-Red requires nodes to be equipped with multiple radios with each radio connected to an independent Wi-Fi networks. 
A link redundancy entity (LRE) is responsible for frame duplication on independent MAC layers, on the transmitter side, and performing duplicate detection on the receiver side. Wi-Red achieves higher reliability through additional hardware complexity. Although Wi-Red adopts PRP, each independent Wi-Fi network still follows the legacy CSMA/CA protocol due to which deterministic latency cannot be guaranteed. Wei \textit{et al.} proposed RT-WiFi \cite{rt_wifi} which aims to provides deterministic timing guarantees on packet delivery while supporting high sampling rates. RT-WiFi implements a TDMA-based MAC layer over the standard 802.11 PHY layer. It incorporates configurable components for various design trade-offs. TDMA-based protocols have also been proposed for multi-hop Wi-Fi networks. Notable studies include Soft-TDMAC \cite{soft_tdmac} and Det-WiFi \cite{det_wifi}. 
The Industrial WLAN (IWLAN) \cite{S_IWLAN}, which is a proprietary technology, addresses the issues of deterministic and reliable performance through PCF at the MAC layer and seamless redundancy using PRP, respectively.
The IEEE 802.11e amendment \cite{802_11e} defines the hybrid coordination function (HCF) which provides two new channel access techniques: enhanced distributed channel access (EDCA) and HCF controlled channel access (HCCA). EDCA differentiates packets using different priorities mapped to specific access categories.
Although EDCA supports traffic prioritization, it does not provide deterministic performance. HCCA improves deterministic performance of PCF by allowing contention-free periods to be initiated anytime during a contention period. However, it has been rarely implemented in conventional WLANs.
Some recent studies like \cite{11ax_RM} and \cite{11ax_S} have  explored resource management aspects of IEEE 802.11ax WLANs. However, the use of IEEE 802.11ax for industrial communication has been rarely investigated. 


\section{\textsf{HAR\(^\text{2}\)D-Fi} -- Design and Protocol Operation} \label{sect_ka}
\subsection{Motivation and Key Aspects}
\textsf{HAR\(^\text{2}\)D-Fi} has been specifically designed for control-centric industrial communication. \textsf{HAR\(^\text{2}\)D-Fi} adopts \textit{hybrid channel access} schemes to provide deterministic communication. Such hybrid channel access is realized by TDMA over DCF, TDMA over DCF with PCF, or window-based DCF schemes.  \textcolor{black}{To achieve high reliability, \textsf{HAR\(^\text{2}\)D-Fi} implements a \textit{temporal redundancy} scheme. The main objective of temporal redundancy is to integrate frequency, time and spatial diversities. Such temporal redundancy provides PRP-like functionality without the hardware complexity of dual radios at each device. To achieve temporal redundancy, \textsf{HAR\(^\text{2}\)D-Fi} exploits  connectivity with two different access points via  \textit{dual association} in a standard-compatible way. } 
The OFDMA-based MAC design has additional distinguishing features. It provides a \textit{virtual TDMA} functionality by exploiting the key capabilities of OFDMA. This is unlike OFDM-based MAC designs which involve some notion of (physical) TDMA that requires a central entity to compute a schedule and allocate timeslots. It also enables \textit{multi-user transmissions} that reduce latency. Besides, it supports \textit{rate adaptation} (for enhancing packet delivery performance in relatively bad channel conditions) by computing an optimal scheduling parameter. 

\textcolor{black}{\textsf{HAR\(^\text{2}\)D-Fi} has been designed for providing reliable and deterministic connectivity over both legacy (OFDM-based) and emerging (OFDMA-based) IEEE 802.11-based chipsets. \textsf{HAR\(^\text{2}\)D-Fi} consists of four different MAC designs. MAC Designs 1 - 3 are based on OFDM-based PHY layer whereas MAC Design 4 is based on an OFDMA-based PHY layer. Each of the MAC Designs 1, 2 and 3 has its own merits from reliability, determinism, efficiency and implementation perspective.  }

\begin{figure}
\centering
\includegraphics[scale=0.19]{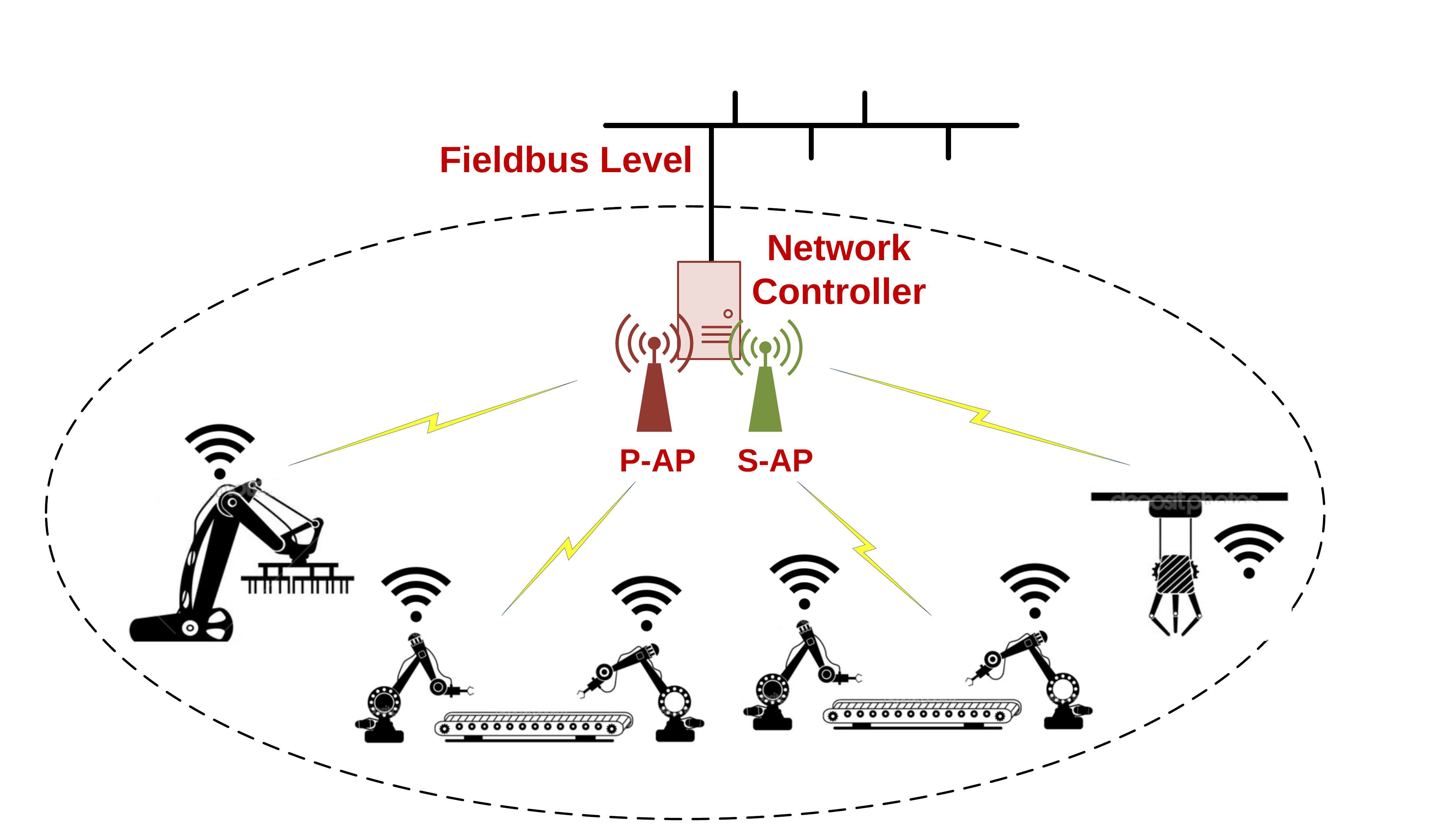}
\caption{An illustration of the system model for \textsf{HAR\(^\text{2}\)D-Fi} operation. }
\label{H_top}
\end{figure}

\subsection{Network Model}
We consider a single-cell scenario, as shown in Fig. \ref{H_top}, wherein wireless connectivity is provided between an industrial controller and multiple field devices like  fixed sensors/actuators or mobile robots which are referred to as stations (STAs). The network controller is split into: (i) an industrial controller which runs the control application, and (ii) a wireless system controller which handles the operation of the wireless network. The \textsf{HAR\(^\text{2}\)D-Fi} system uses two access points: a primary access point (P-AP) and a secondary access point (S-AP). Both  access points are connected to the wireless system controller through a wired interface (e.g., Ethernet). The P-AP and the S-AP have different service set identifiers (SSIDs) and operate on different channels while providing overlapping coverage. Time synchronization  is achieved through the IEEE 802.11 timing synchronization function (TSF) wherein STAs maintain a local \(1\) MHz timer which is periodically synchronized through beacons received from the access point. 
\textsf{HAR\(^\text{2}\)D-Fi} exploits two different PHY layers. The first PHY layer is based on OFDM as per the legacy IEEE 802.11 specifications. The MAC layer in this case is agnostic to the underlying 802.11 standard (.11a, .11n, .11ac, etc.). The second PHY layer is based on OFDMA, as per the IEEE 802.11ax specifications, which divides a Wi-Fi channel into smaller sub-channels. The .11ax PHY layer has the same channel configuration as .11ac. Additionally, it supports \(2.22\) MHz, \(5\) MHz, and \(10\) MHz sub-channel widths. The notion of a timeslot is common to all \textsf{HAR\(^\text{2}\)D-Fi} MAC designs and dependent on the net payload and the PHY layer data rate. For example, a \(128\) byte packet would need \(1024\) \(\mu\)s and  \(19\) \(\mu\)s at \(1\) Mbps  and \(54\) Mbps, respectively. 



\subsection{MAC Design 1: TDMA over DCF}
\begin{figure}
\centering
\includegraphics[scale=0.24]{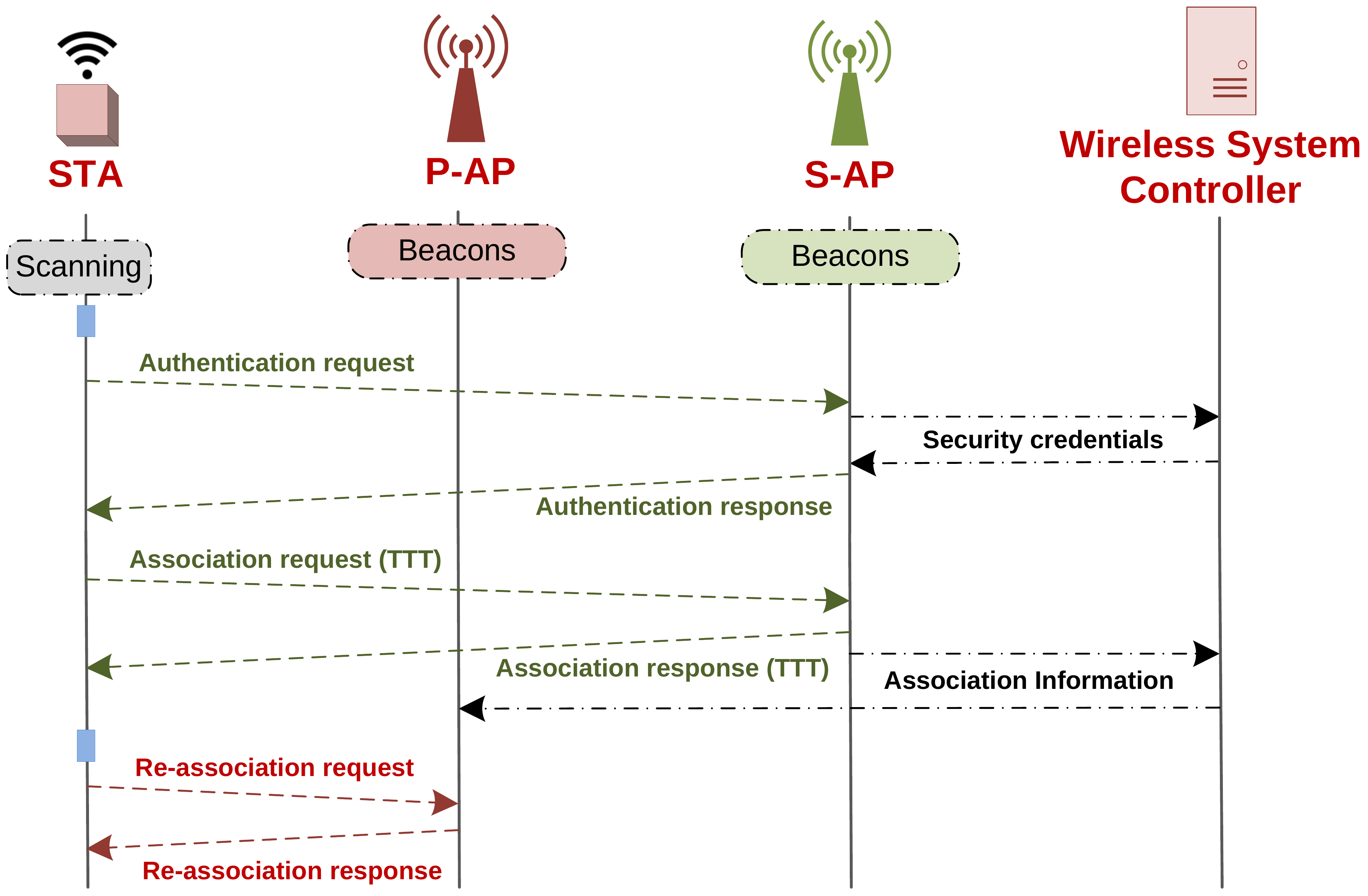}
\caption{The dual association method in \textsf{HAR\(^\text{2}\)D-Fi}. }
\label{H_dual_assoc}
\end{figure}

The first MAC layer design adopts TDMA over conventional DCF-based channel access. TDMA provides a centralized medium access according to a schedule such that only one node can access the channel in a given timeslot. The wireless system controller is responsible for computing a schedule according to any pre-defined criteria, for example based on a round robin approach.  To exploit temporal redundancy, STAs must  have \textit{a priori} association with two access points. However, the legacy 802.11 protocol does not support this functionality. Therefore, we first propose a standard-compatible dual association method which is illustrated in Fig. \ref{H_dual_assoc} and described as follows.

In 802.11-based WLANs, data transfer between a STA and an access point cannot take place unless the former is in \textit{authenticated and associated} state. Initially, both access points transmit beacon frames on their respective channels. The beacons also contain information about the channel used by the other access point. The STAs, on initial power up,  actively or passively scan the channels for beacons.  The S-AP has priority over the P-AP during initial network association. 
If a STA receives beacons from the P-AP, it will switch channel to that of the S-AP. A STA proceeds to the authentication
phase after it receives beacons from the S-AP. It sends an authentication request frame to the S-AP. The S-AP performs a security check with the wireless system controller and responds  with an authentication response frame. After authentication, the respective STA sends an association request frame to the S-AP which contains an important parameter, called \textit{target transmission time} (TTT), that indicates the minimum duration for which the STA will not transmit or receive from the S-AP for the very first time. The computation of TTT is discussed later. The S-AP responds with an association response frame confirming successful negotiation of the TTT parameter. Hence, the respective STA is successfully associated with the S-AP. The S-AP shares the association information with the P-AP through the wireless system controller. Therefore, the P-AP has \textit{a priori} knowledge of  association of a STA with the S-AP. Once a STA is associated to the S-AP, it sends a re-association request frame to the P-AP. Since the P-AP is aware of the authentication of the respective STA, there is no need for  re-authentication. The P-AP responds with a re-association response frame to the respective STA. Hence, a STA is successfully associated to two access points. Note that a STA will only transmit and receive data from a single access point at any given time. \textcolor{black}{Note that dual association is only required before the start of communication cycle between the controller and the devices. Hence, its delay and overhead is not critical.}

In \textsf{HAR\(^\text{2}\)D-Fi} the IP address assignment functionality is decoupled from the access points and realized through the wireless system controller. A STA is only assigned an IP address  during association  with the S-AP. 
Note that IP address assignment may not be necessary as the application layer directly sits on top of the MAC layer in most industrial applications \cite{w_low_fa}.

\begin{figure}
\centering
\includegraphics[scale=0.21]{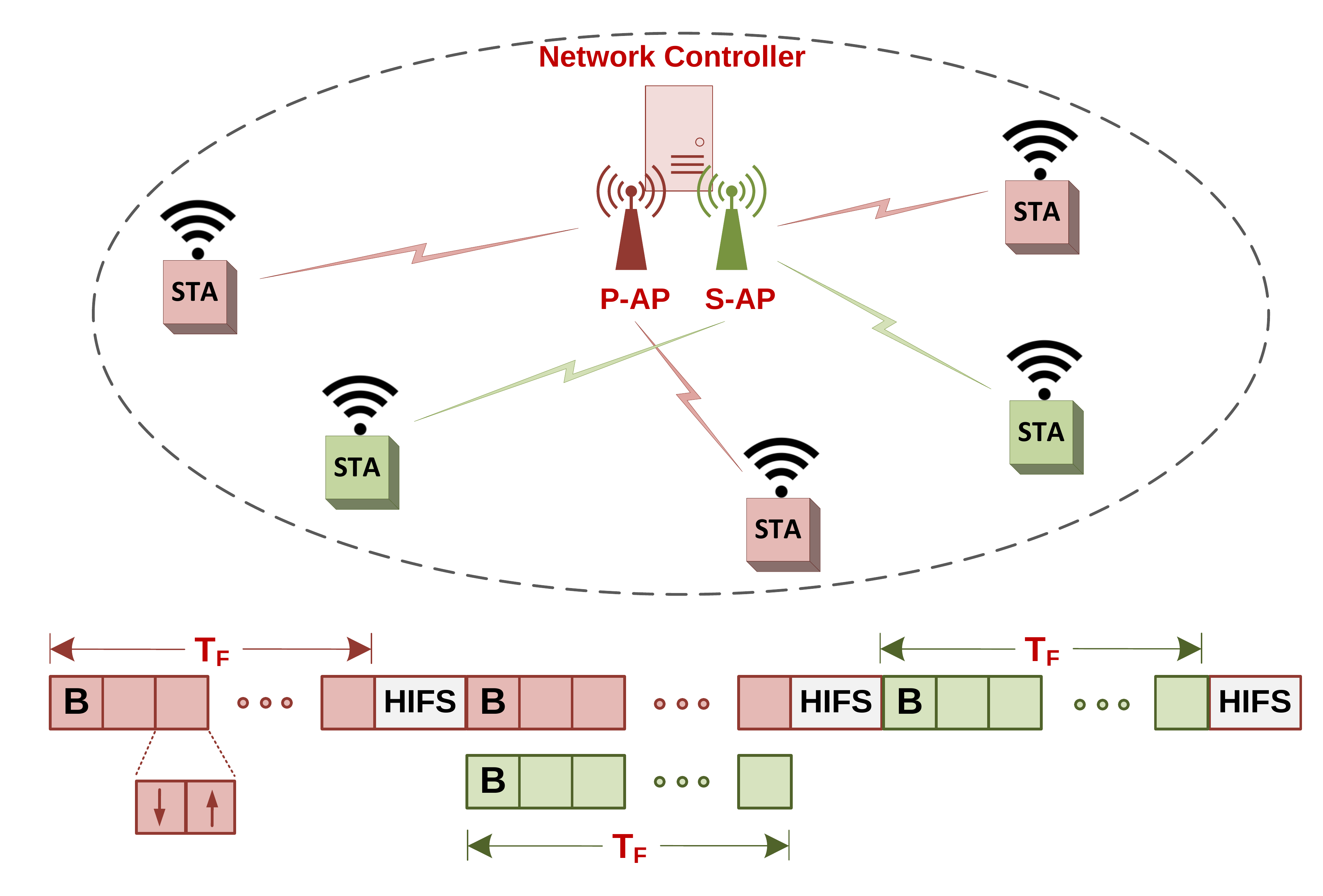}
\caption{Operation of \textsf{HAR\(^\text{2}\)D-Fi} MAC Design 1 with two groups. Red and green refer to operation with the P-AP and the S-AP, respectively.}
\label{HFi_MD1}
\end{figure}


Next, we explain the MAC design.   Let, \(N\) denote the number of STAs within the coverage of the two access points. After the association phase, the wireless controller splits these STAs into \(M\) groups (with distinct group IDs). The MAC layer is based on a superframe structure such that it consists of \(M+1\) frames, with each frame comprising \(N/2+1\) double timeslots. Each double timeslots consists of two single timeslots which are used for bidirectional\footnote{Typically, control applications involve bidirectional cyclic information exchange between a controllers and devices.  The period of control cycle is referred to as the \textit{cycle time} which is defined as  the total time it takes from successfully transmitting command messages to all the devices to successfully receiving the corresponding response message from each device.  } communication between the access point and the STA, respectively.  The scenario of two groups is depicted in Fig. \ref{HFi_MD1}. The first double timeslot of each frame is reserved for beacon transmissions from the access point which provide time synchronization. These transmissions also include schedule information for the STAs which consists of a pair of group ID and allocated timeslot within the frame. Two adjacent frames are separated by \textsf{HAR\(^\text{2}\)D-Fi} interframe space (HIFS) duration. Initially, the first group of STAs communicate with the P-AP over the first frame. Based on the schedule information, each double timeslot is allocated to a specific STA. The first timeslot in a double timeslot is allocated for downlink transmission from the AP whereas the second is allocated for uplink transmission from the STA. The duration of each single timeslot is given by 
\begin{equation}
\label{ts_dur}
T_{\text{slot}}=T_{\text{Data}}+T_{\text{SIFS}}+T_{\text{ACK}}+T_{\text{guard}},
\end{equation}
where \(T_{\text{Data}}\) and \(T_{\text{ACK}}\) denote the transmission time for data and acknowledgement (ACK) transmissions, and \(T_{\text{SIFS}}\) and \(T_{\text{guard}}\) denote the short interframe space (SIFS) and guard interval duration, respectively. 
The bidirectional communication in a double timeslot can be optimized with a bidirectional transmit (BDT) feature where a STA transmits  data instead of  an ACK. 

After finishing communication with the P-AP, the first group of STAs perform a channel switching operation to that of the S-AP. The HIFS duration accounts for the channel switching time which is dependent on the  hardware. The first group of STAs follow a similar procedure as before, and communicate with the S-AP. The first double timeslot is reserved for beacon transmissions from the S-AP which provide time synchronization and schedule information. Note that the transmission of schedule information in this case is not mandatory as it is computed by the wireless controller and already received by the STAs from the P-AP. However, retransmission of the schedule information provides redundancy which is particularly attractive from reliability perspective. It is emphasized that communication with the S-AP provides \(100\%\) redundancy as transmissions are repeated irrespective of the success or failure with the P-AP. 

Whilst the first group of STAs is communicating with the S-AP, the second group communicates with the P-AP. After the end of the second frame, the second group of STAs communicates with the S-AP for repeating transmissions. It is noteworthy that without the dual association method, there would be a non-deterministic delay to account for association of a group of STAs with the respective AP after every frame. The requirement to repeat transmissions subsequently on the S-AP,  exactly as they are carried out on the P-AP, not only provides temporal redundancy but also greatly simplifies implementation. Such temporal redundancy provides a PRP-like functionality without complexity. The superframe structure can be repeated to account for any transmission failures depending on the latency and reliability requirements.

\textcolor{black}{The duration in TTT parameter depends on the number of groups, the specific group to which the STA belongs and the superframe length. However, no prior knowledge of grouping is available to the STAs during association. A STA sets the TTT parameter to one frame duration initially. The wireless controller has  knowledge of the number of groups. Therefore, during the association phase, the S-AP can update the TTT based on the information obtained from the wireless controller.}


\begin{figure}
\centering
\includegraphics[scale=0.165]{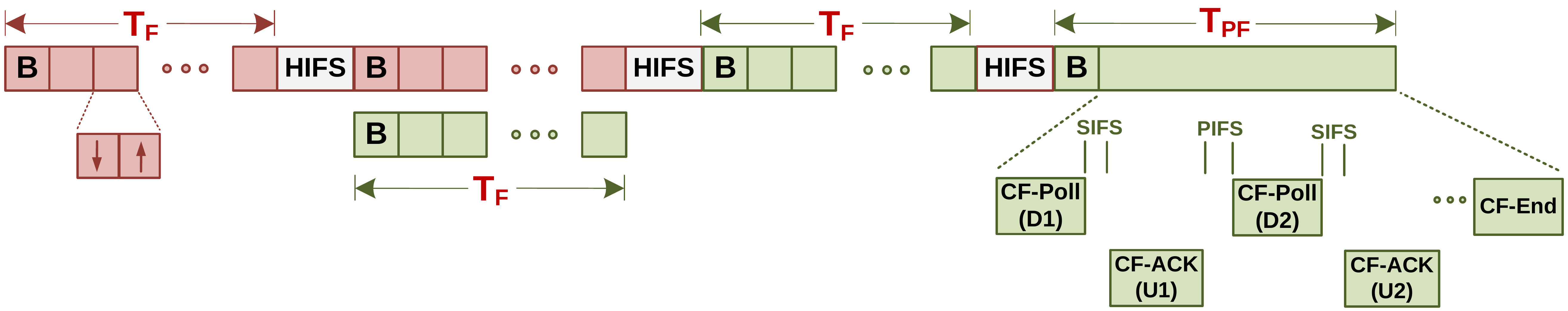}
\caption{Operation of \textsf{HAR\(^\text{2}\)D-Fi} MAC Design 2.  }
\label{HFi_MD2}
\end{figure}

\subsection{MAC Design 2: TDMA over DCF with PCF}
The second MAC design improves the efficiency of MAC Design 1 through a PCF-based functionality. The proposed MAC design is illustrated in Fig. \ref{HFi_MD2} and explained as follows. 

The MAC Design 2 is split into two phases. The first phase follows the same operation as MAC Design 1. The second phase of MAC Design 2 adopts PCF-based operation for handling retransmissions arising from failures in the first phase. A transmission failure can occur if either uplink or downlink transmissions are not successful with both the P-AP and the S-AP. Unlike MAC Design 1 where the superframe is repeated for handling retransmissions, the approach herein is to selectively repeat transmissions from the unsuccessful nodes. The second phase consists of a single or multiple PCF frames, each of which starts with beacon transmissions from the access point that signal the beginning of polling phase and contain information about the STAs requiring retransmissions. After transmitting the beacons, the access point starts polling the unsuccessful STAs in a round robin fashion. The access point transmits a CF-Poll message to a STA. As the access point has no \textit{a priori} knowledge of the cause for failure, it piggybacks the downlink information for the respective STA on the CF-Poll message. After receiving the CF-Poll message, the STA waits for SIFS duration and responds with a CF-ACK message which contains the uplink message for the access point. The access point repeats the operation with the next unsuccessful STA after waiting for PCF interframe space duration (PIFS). The polling phase continues until all retransmissions are successful. At the end of the polling phase, the access point transmits a CF-End signal to let all STAs know that polling phase has ended and that they should now tune to the P-AP to start the next cycle of transmission. 

The second phase of MAC Design 2 can be handled through the P-AP or the S-AP. However, this has to be fixed at design time.
 The second phase can also be handled through both the access points, i.e., with separate PCF frames.

\begin{figure}
\centering
\includegraphics[scale=0.16]{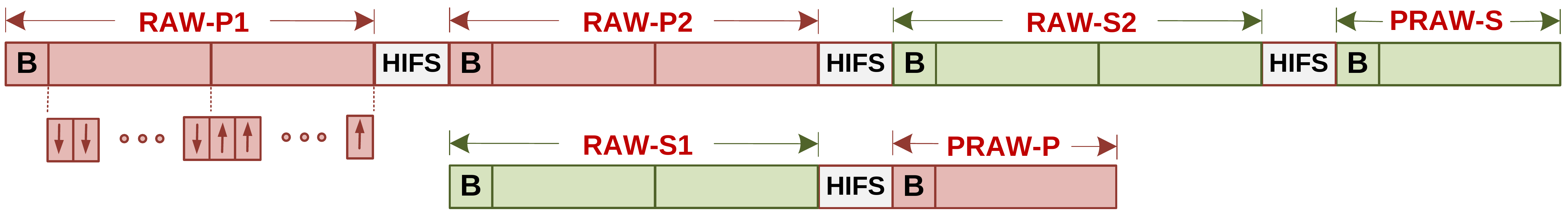}
\caption{Operation of \textsf{HAR\(^\text{2}\)D-Fi} MAC Design 3.  }
\label{HFi_MD3}
\end{figure}

\subsection{MAC Design 3: Window-based DCF}
The third MAC design is based on the concept of restricted access window (RAW) which was first proposed in the IEEE 802.11ah-2016 standard \cite{802_11_ah}. The fundamental principle of RAW is to restrict channel access only to a given group of STAs. In MAC Design 3, each RAW handles bidirectional communication for a fixed number of STAs. For instance, in case of two groups, each RAW handles the operation of \(N/2\) STAs. The first two timeslots of each RAW are reserved for beacon transmissions. Each RAW consists of downlink and uplink subframes which are further split into a fixed number of timeslots. The beacon transmissions contain a traffic information map (TIM) \cite{802_11_ah} that provides timeslot allocation information to STAs, i.e., downlink and uplink timeslot in the respective subframe. The access point transmits control messages to different STAs in their allocated timeslots of the downlink subframe. The respective STA responds in the allocated timeslot of the uplink subframe. 
Initially, the RAWs are advertised by the P-AP. The first group of STAs communicates with the P-AP as shown in Fig. \ref{HFi_MD3}. Once the first group of STAs finishes communication with the P-AP, it performs a channel switching operation to that of the S-AP.  The STAs follow a similar approach as before and communicate with the S-AP based on the received TIM.  While the first group of STAs is engaged in communication with the S-AP, the second group communicates with the P-AP. In the third frame, the second group of STAs communicates with the S-AP. 

\textcolor{black}{Unlike MAC Design 1 and 2, MAC Design 3 has  additional provisioning for uplink only traffic. This is realized through a periodic RAW (PRAW) which contains a fixed number of timeslots for uplink only traffic. The superframe structure in this case consists of \(M+1\) RAWs and \(1\) PRAW. Provisioning for uplink traffic is an optional feature in MAC Design 3. Finally, the MAC layer retransmissions in this case  are handled through the PCF-based operation as described for MAC Design 2.}

\subsection{MAC Design 4: Virtual TDMA}
The fourth MAC design exploits the key capabilities of the next generation Wi-Fi standard in providing a virtual TDMA functionality. It enables multi-user transmissions by allocating non-overlapping sub-channels to multiple users which transmit in parallel. Such multi-user transmissions are initiated through trigger frames. 
The MAC layer design  is based on different types of frames. Unlike previous designs, there is no pre-defined allocation of timeslots for such frames. Let, \(W_{\text{ch}}\) and \(W_\text{s}\) denote the Wi-Fi channel bandwidth and the sub-channel bandwidth, respectively. The number of devices which can be simultaneously supported is given by \(W_{\text{ch}}/W_\text{s}\). For instance, with \(W_{\text{ch}}\) = \(20\) MHz and  \(W_\text{s}\) = \(2.22\) MHz (minimum supported sub-channel), up to \(9\) devices can be simultaneously supported.

\begin{figure}
\centering
\includegraphics[scale=0.26]{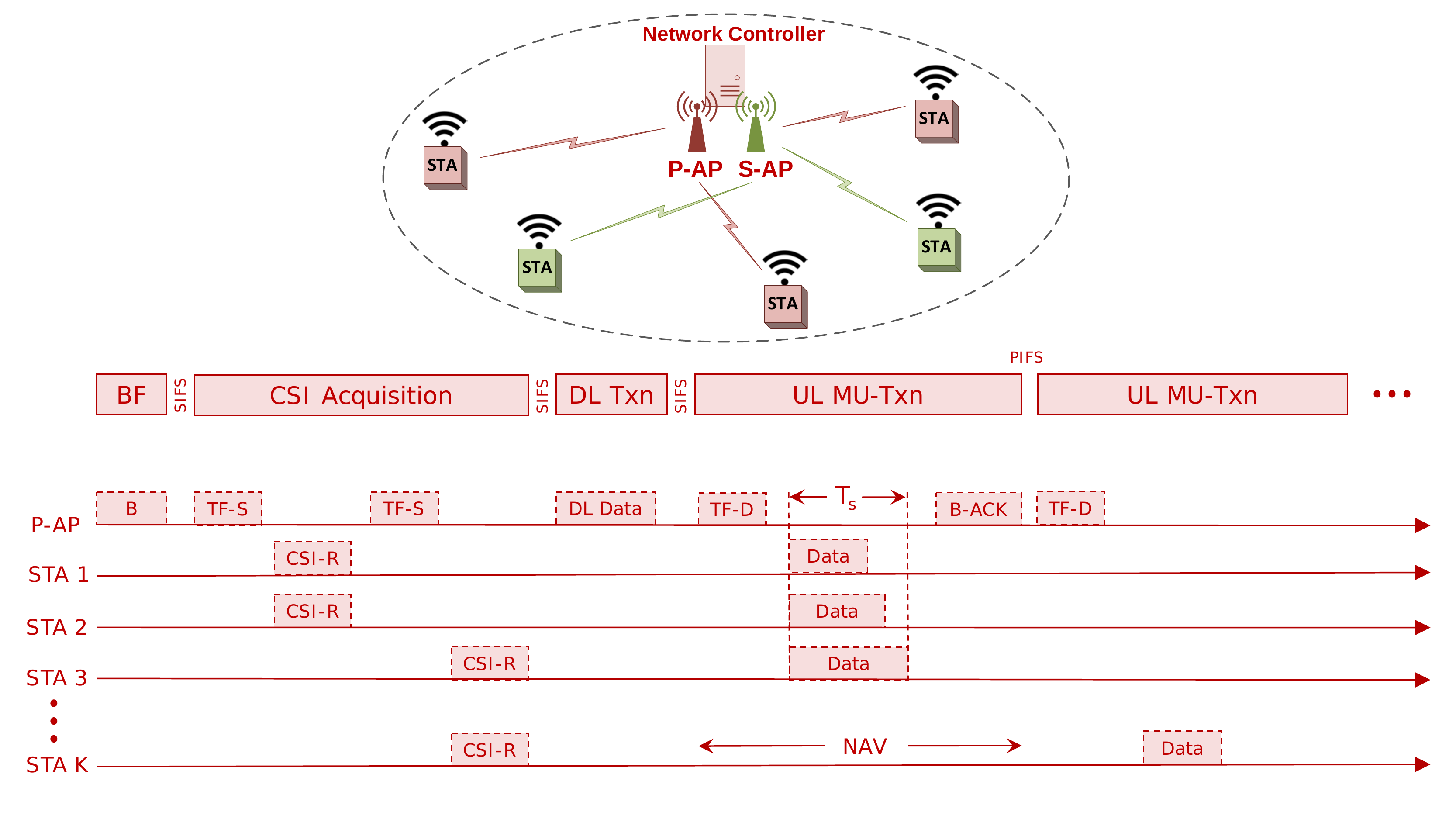}
\caption{ \textsf{HAR\(^\text{2}\)D-Fi} MAC Design 4: MAC layer operation for the P-AP}
\label{HFi_MD4}
\end{figure}

\begin{figure}
\centering
\includegraphics[scale=0.26]{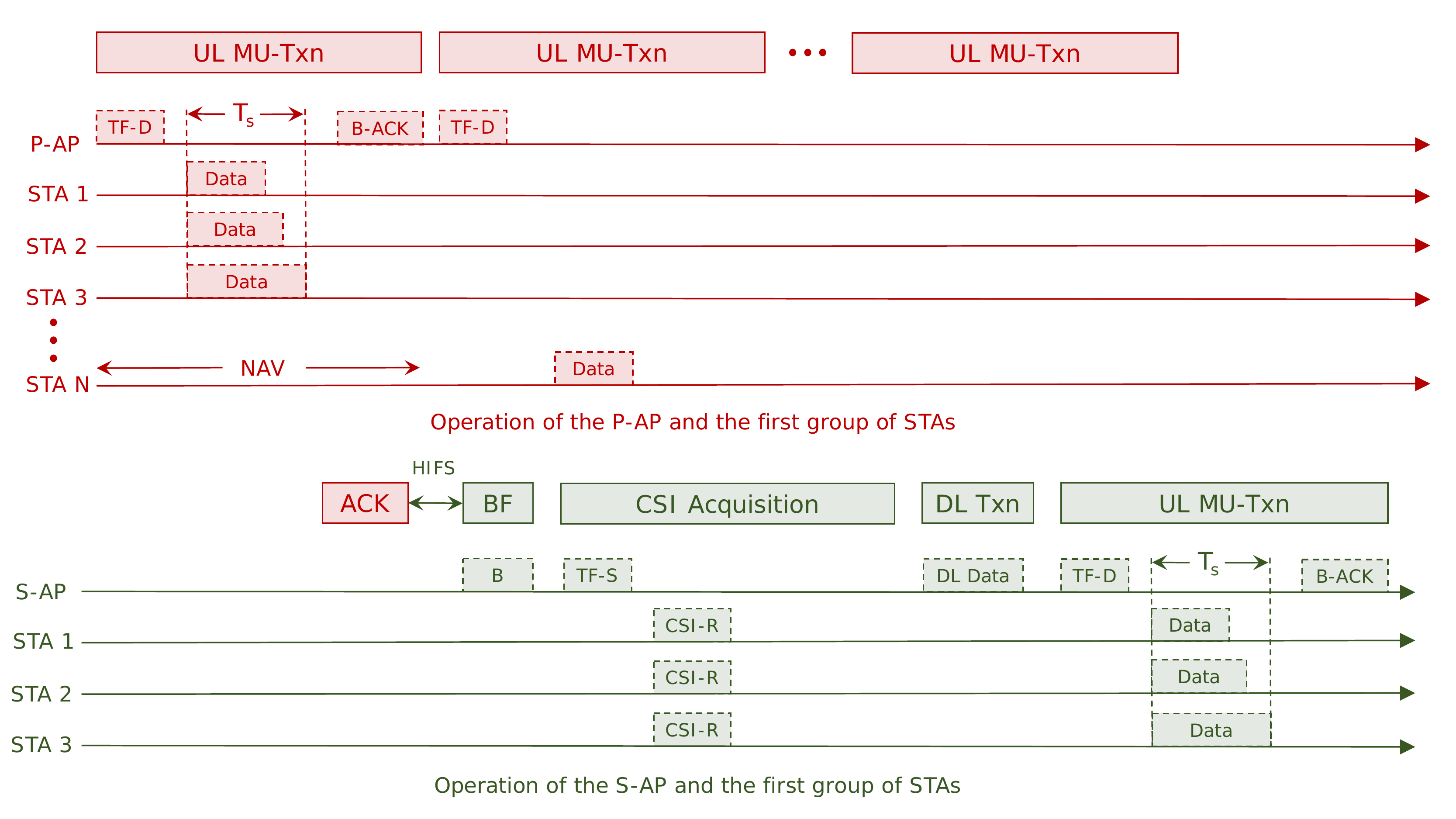}
\caption{\textsf{HAR\(^\text{2}\)D-Fi} MAC Design 4: MAC layer operation of the P-AP and the S-AP with temporal redundancy.}
\label{HFi_MD4b}
\end{figure}


We explain the protocol operation with the aid of Fig. \ref{HFi_MD4} and Fig. \ref{HFi_MD4b}  which show the timeline of both access points. The protocol initiates with a beacon frame which contains beacon transmissions carrying network and synchronization information. In order to realize the rate adaptation functionality, the access point must acquire the channel state information (CSI) from the STAs in the network. The frequency of obtaining CSI is dependent on the nature of the environment. 
 Therefore, the beacon frame is followed by a CSI acquisition frame. After waiting for a SIFS duration, the P-AP transmits a trigger frame which solicits CSI from a group of STAs through multi-user transmissions. It contains information about the STAs involved in the multi-user transmission along with the sub-channel allocation information. In response to the trigger frame, STAs send their CSI reports after waiting for SIFS duration. Note that multiple trigger rounds might be needed to gather information from all the STAs in the network. After acquiring the CSI, the P-AP proceeds to scheduling downlink and uplink transmissions. For most industrial control applications the cycle starts with downlink commands from the controller. Therefore, the CSI acquisition frame is followed by the downlink transmission frame. The downlink transmission by the P-AP takes place after a SIFS duration from end of the CSI Acquisition phase.  Similar to beacon transmissions, the downlink transmissions can be sent as a broadcast. However, the P-AP can also perform multi-user downlink transmissions. Note that a trigger frame is not required for downlink transmission as it is initiated by the P-AP. The downlink transmission frame is followed by single or multiple uplink transmission frames. After waiting for a SIFS duration from the end of the downlink transmission, the P-AP transmits a trigger frame for scheduling multi-user uplink transmission. Note that this trigger frame is different than the one sent for acquiring CSI. The trigger frame contains information about the users involved in the multi-user transmission along with user-specific information including the allocation of sub-channels and modulation and coding scheme. In response to the trigger frame, the corresponding STAs perform data transmission in the uplink. The multi-user transmission must finish at the same time for all STAs. In this context, the trigger frame also contains an optimal scheduling duration parameter denoted by \(T_\text{s}\). The STAs with transmission duration less than \(T_\text{s}\) append some padding bits to end the transmission at the same time as other STAs. Note that the variation in transmission duration occurs due to the aforementioned rate adaptation functionality. After waiting for a SIFS duration from \(T_\text{s}\) the P-AP transmits a block ACK (single ACK for all transmissions) to all the STAs engaged in multi-user uplink transmission.  Note that the trigger frames and block ACKs are transmitted over the entire channel bandwidth. 

\begin{algorithm}
 \caption{\textsc{\textcolor{black}{\(T_s\) Calculation (Fixed Payload)}}}
\textbf{Input}: \(L\) (net payload for uplink transmission), \\
\quad \(K_s\) (set of STAs for multi-user transmission), \\
\quad  \(C_s\) (set of CSI reports), 
\  \(\mathcal{Q}\) (CSI mapping table) \\ 
\textbf{Step} $1$:
find STA \(k^*\) with lowest SNR, \(S_{k^*}\) \\
\textbf{Step} $2$:
map \(S_{k^*}\) to appropriate MCS in \(\mathcal{Q}\) \\
\textbf{Step} $3$:
calculate \(R_{k^*}=W_s \log_2(1+ S_{k^*})\)\\
\textbf{Step} $4$:
calculate \(T_s=\lceil L/R_{k^*} \rceil\)\\
\label{algo_ts1}
\end{algorithm}

Note that the STAs which are not part of the multi-user transmission set their network allocation vector (NAV) to the duration advertised in the trigger frame and defer from accessing the medium. To further protect uplink transmissions, the P-AP can  utilize a multi-user request-to-send/clear-to-send (RTS/CTS) method. Note that multiple multi-user uplink transmissions might be needed as part of a control cycle. Each multi-user uplink transmission is separated by a PIFS duration which ensures that the P-AP retains priority in channel access. 

Next, we explain how temporal redundancy is achieved in this design. After finishing the first multi-user uplink transmission, the P-AP exchanges the list of successful STAs with the S-AP through the wireless controller. The group of STAs which were part of the first multi-user uplink transmission perform a channel switching information to that of the S-AP. The operation of S-AP starts after a HIFS duration from the end of first multi-user uplink transmission. The S-AP performs a similar operation as the P-AP, i.e., it transmits beacons, acquires CSI, performs downlink transmission, and schedules multi-user uplink transmissions. However, it only engages in transmission with the STAs which have finished their multi-user transmission with the P-AP.  The detailed operation of the S-AP is shown in Fig. \ref{HFi_MD4}. Finally,  MAC layer retransmissions can be handled through trigger frames and multi-user transmissions by one of the access points.

Next, we discuss the calculation of the optimal scheduling time, \(T_s\), which is illustrated by Algorithm \ref{algo_ts1}. The CSI report from a  STA provides channel quality information, e.g., in terms of received signal-to-noise ratio (SNR), to the access point. In practice, the access point only supports a discrete  set of  modulation and coding schemes (MCSs).   Based on  CSI reports, the access point selects an appropriate MCS for each STA, in order to perform rate adaptation. 
For the sake of calculating \(T_s\), the access point selects STA \(k^{*}\) with the lowest SNR 
as shown in Algorithm \ref{algo_ts1}.  For most control applications, the payload is typically fixed. The proposed algorithm can be  extended to account for variable payload per transmission 
as shown in Algorithm \ref{algo_ts2}.
In this case, the access point calculates the transmission time for each STA, depending on the assigned MCS and the uplink payload, and finds the optimal scheduling time as the maximum of the calculated values.

\begin{algorithm}
 \caption{\textsc{\textcolor{black}{\(T_s\) Calculation (Variable Payload)}}}
\textbf{Input}: \(L_s\) (set of net uplink payload for each STA) \\
\quad \qquad \(K_s\); \(C_s\);  \(\mathcal{Q}\) \\
\textbf{Step} $1$:
find SNR for each STA, \(S_i\) \\
\textbf{Step} $2$:
map SNR to appropriate MCS in \(\mathcal{Q}\) for each STA \\
\textbf{Step} $3$:
calculate \(R_{i}=W_s \log_2(1+ S_{i})\)\\
\textbf{Step} $4$:
calculate \(T_s=\max (\lceil L_i/R_{i} \rceil); \ i \in L_s\)\\
\label{algo_ts2}
\end{algorithm}

\section{\textsf{HAR\(^\text{2}\)D-Fi} -- Analytic Framework} \label{sect_af}
In this section we develop an analytic framework for evaluation of \textsf{HAR\(^\text{2}\)D-Fi}. \textcolor{black}{The aim of the analytic framework is to integrate spatial dynamics with protocol aspect in order to derive closed-form expressions for different performance metrics.}  We consider a single-cell scenario wherein the P-AP and the S-AP are co-located and provide overlapping coverage with radius \(\mathcal{R}_{A}\).  We adopt the standard signal-to-noise ratio (SNR) model for link outages. Based on this model, a transmission is successful if the instantaneous SNR exceeds a certain threshold. 
Let \(\gamma_{\text{DL}}\) denote the instantaneous downlink SNR at a STA, such that \(\gamma_{\text{DL}}=\mathcal{P}^t_A\vert h \vert^2 r^{-\alpha}/\sigma^2\), where \(\mathcal{P}^t_A\) is the transmit power of the access point, \(h\) is the channel fading coefficient, \(r\) is the distance between the STA and any of the access points, \(\alpha\) is the path loss exponent, and \(\sigma^2\) denotes the noise power. 
Let, \(P_{\text{DL}}^o\) denote the instantaneous outage probability for a transmission in the downlink, which can be calculated as 
\begin{equation}
\label{dl_op_1a}
P_{\text{DL}}^o=\mathbb{P} \left\lbrace \gamma_{\text{DL}} \leq \beta \right\rbrace=\mathbb{P} \left\lbrace \vert h \vert^2 \leq \beta \sigma^2 r^\alpha/\mathcal{P}^t_A \right\rbrace,
\end{equation}
where \(\beta\) is the threshold SNR for successful transmission. In case of widely used Rayleigh fading, the probability density function (PDF) of  \(\vert h \vert^2\) is exponentially distributed with mean \(1/ \mu\), and the instantaneous outage probability is given by 
\begin{equation}
\label{op_dl2}
P_{\text{DL}}^o=1-\exp \left(-\frac{\mu \beta \sigma^2 r^\alpha}{\mathcal{P}^t_A} \right).
\end{equation}

We assume that \(N\) STAs are \textit{uniformly} distributed within \(\mathcal{R}_{A}\). The PDF of the distance \(r\) between any of the access points and the \(n\)th neighboring STA is given by \cite{dist_reg}
\begin{equation}
\label{pdf_dist_1}
f(r,n)=\frac{\left(1-r^2/\mathcal{R}_{A}^2\right)^{N-n} \left( r^2/\mathcal{R}_{A}^2\right)^{n-1}}{B\left( N-n+1, n\right)} \frac{2r}{\mathcal{R}_{A}^2},
\end{equation}
where \(B(.)\) is the Beta function.
Using  \eqref{op_dl2} and \eqref{pdf_dist_1}, the average outage probability in the downlink is given by
\begin{equation}
\label{o_dl_a}
\begin{aligned}
\allowdisplaybreaks
\bar{P}_{\text{DL}}^o=\mathbb{E}_r \left\lbrace P_{\text{DL}}^o \right\rbrace
&=\frac{1}{N}\sum_{k=1}^N\int_0^{\mathcal{R}_{A}}\left[ 1-\exp \left(-\zeta_{d}  r^\alpha \right)  \right] f(r,k) \mathrm{d}r \\ 
&=\frac{2}{\mathcal{R}_{A}^2}\int_0^{\mathcal{R}_{A}} r \left[ 1-\exp \left(-\zeta_{d}  r^\alpha \right)  \right] \mathrm{d}r, 
\end{aligned}
\end{equation}
where \(\zeta_{d}= \mu \beta \sigma^2/\mathcal{P}_{A}^t\). The average outage probability in the uplink can be obtained in a similar way as \eqref{o_dl_a} with \(\zeta_{d}\) replaced by \(\zeta_{u}= \mu \beta \sigma^2/\mathcal{P}_{S}^t\) such that \(\mathcal{P}_{S}^t\) denotes the transmit power of the STA.  
A transmission with the P-AP or the S-AP fails if it is unsuccessful in either uplink or downlink. The outage probability for a transmission with the P-AP  can be calculated as \(P_{\text{P-AP}}^{o}=\bar{P}_{\text{DL}}^o+\bar{P}_{\text{UL}}^o\).
The outage probability for a transmission with S-AP, i.e., \(P_{\text{S-AP}}^{o}\) can be obtained in a similar way. Note that \(\bar{P}_{\text{DL}}^o\) accounts for the transmit power of the P-AP or the S-AP, denoted by \(\mathcal{P}_{\text{P-AP}}^t\) and \(\mathcal{P}_{\text{S-AP}}^t\), respectively. 

A transmission failure occurs if a transmission is unsuccessful with both access points, i.e., \(P_{\text{fail}}=P_{\text{P-AP}}^{o} \cdot P_{\text{S-AP}}^{o}\). Hence, the average number of transmission failures is given by
\begin{equation}
\label{txn_fail}
\Theta_r=\sum_{i=1}^N i {N \choose i}   \left(P_{\text{fail}}\right)^i \left(1-P_{\text{fail}} \right)^{N-i} = N \cdot P_{\text{fail}}.
\end{equation}

Next, we calculate the cycle time for different MAC designs. In case of MAC Design 1, the cycle time is given by
\begin{equation}
\label{cyc_MD1}
\mathcal{C}_1=\left(M+1\right)\left(T_{\text{B}}+\frac{2NT_{\text{slot}}}{M} \right)N_{\text{SF}}+N_{\text{SF}} \cdot M \cdot T_{\text{HIFS}},
\end{equation}
where \(T_{\text{B}}\) is the beacon duration, \(N_{\text{SF}}\) is the number of superframes, \(T_{\text{HIFS}}\) is the HIFS duration, and \(T_{\text{slot}}\) is given by \eqref{ts_dur}.
For MAC Design 2, the first phase is similar to MAC Design 1, whereas the second phase adopts a PCF-based operation. 
The duration of the first PCF frame, which is a function of the number of transmission failures in the first phase, is given by
\begin{equation}
\label{pcf_dur}
T_{\text{PF}}=T_{\text{B}}+ \Theta_r \left(T_{\text{P}}+T_{\text{SIFS}}+T_{\text{A}} \right)+\left(\Theta_r-1 \right)T_{\text{PIFS}}+T_{\text{E}},
\end{equation}
where \(T_{\text{P}}\), \(T_{\text{A}}\) and \(T_{\text{E}}\) denote the duration for CF-Poll, CF-ACK and CF-End respectively, and \(T_{\text{PIFS}}\) is the PIFS duration. 
The average number of transmission failures in the second phase is given by
\begin{equation}
\label{pcf_no}
\Theta_{p,2}=\sum_{i=1}^{\Theta_r} i {\Theta_r \choose i}   \left(P_{\text{fail}}\right)^i \left(1-P_{\text{fail}} \right)^{\Theta_r-i} = \Theta_r \cdot P_{\text{fail}}.
\end{equation}
Hence, the cycle time for MAC Design 2 can be computed as 
\begin{equation}
\label{cyc_MD2}
\mathcal{C}_2=\mathcal{C}_1\vert_{N_{\text{SF}}=1}+T_{\text{PF}}+\sum_{j=2}^{H}  \Theta_{p,j} \cdot   T_{\text{PF}}\vert_{\Theta_r=\Theta_{p,j}},
\end{equation}
where \(H\) denotes the maximum number of retransmission phases, and \(\mathcal{C}_1\) and \(T_{\text{PF}}\) are given by \eqref{cyc_MD1} and \eqref{pcf_dur}, respectively. 

The duration of a single RAW for MAC Design 3 is given by \(T_{\text{RAW}}=T_{\text{B}}+2T_{\text{slot}}N/M\). Hence, the cycle time is given by \eqref{cyc_MD3}, where \(T_{\text{PRAW}}\) is the PRAW duration. 
\begin{equation}
\begin{aligned}
\label{cyc_MD3}
\mathcal{C}_3=\left(M+1\right)T_{\text{RAW}}+T_{\text{PRAW}}+T_{\text{PF}}
+\sum_{j=2}^{H}  \Theta_{p,j} \cdot   T_{\text{PF}}\vert_{\Theta_r=\Theta_{p,j}},
\end{aligned}
\end{equation}

Next, we analyze the MAC Design 4. For the sake of analysis, we assume that multi-user transmissions are used in both downlink and uplink. The average outage probability for the downlink can be calculated as 
\begin{equation}
\label{o_dl_a_4}
\bar{P}_{\text{DL,4}}^o=1-\left(1-\bar{P}_{\text{DL}}^o\right)^{K},
\end{equation}
where \(K=W_{\text{ch}}/W_\text{s}\) denotes the number of STAs in a multi-user transmission and \(\bar{P}_{\text{DL}}^o\) is given by \eqref{o_dl_a}. The outage probability for the uplink, i.e., \(\bar{P}_{\text{UL,4}}^o\)  can be calculated in a similar way. 
The outage probability for a transmission with the P-AP (denoted by \(P_{\text{P-AP,4}}^{o}\)) or the S-AP (denoted by \(P_{\text{S-AP,4}}^{o}\)) can be computed in a similar way as described for previous MAC designs. The average number of transmission failures in this case is given by
\begin{equation}
\label{txn_fail_4}
\Theta_r^4=\sum_{i=1}^{N^{'}} i {{N^{'}} \choose i}   \left(P_{\text{fail,4}}\right)^i \left(1-P_{\text{fail,4}} \right)^{N^{'}-i} = N^{'} \cdot P_{\text{fail}},
\end{equation}
where \(N^{'}=\lceil N/K \rceil\) and \(P_{\text{fail},4}=P_{\text{P-AP,4}}^{o} \cdot P_{\text{S-AP,4}}^{o}\).
The cycle time for MAC Design 4 depends on different components. The duration of a single multi-user uplink transmission is given by
\(T_{\text{MU-UL}}=T_{\text{TF-D}}+2T_{\text{SIFS}}+T_s+T_{\text{B-ACK}}\),
where \(T_{\text{TF-D}}\) and \(T_{\text{B-ACK}}\) denote the duration of the trigger frame (for data transmission) and the block ACK, respectively. 
The duration of the CSI acquisition phase is given by 
\(T_{\text{CSI}}=N^{'} \left(T_{\text{TF-S}}+T_{\text{CSI-R}}\right)+(2N^{'}-1 )T_{\text{SIFS}}\),
where \(T_{\text{TF-S}}\) and \(T_{\text{CSI-R}}\) denote the duration of the trigger frame (for CSI acquisition) and the CSI report, respectively. 
The cycle time for MAC Design 4 without any MAC layer retransmissions can be calculated as 
\begin{equation}
\begin{aligned}
\allowdisplaybreaks
\label{cyc_MD4}
\mathcal{C}_4^{'}=&T_B+T_{\text{CSI}}+(N^{'}+1)(T_s+T_{\text{MU-UL}})+\\
&2(N^{'}+2)T_{\text{SIFS}}+2(N^{'}-1)T_{\text{PIFS}},
\end{aligned}
\end{equation}
where \(T_{\text{PIFS}}\) is the PIFS duration. The duration of one retransmission phase is given by 
\(T_{\text{Ret}}=\mathcal{C}_4^{'}\vert_{N^{'}=\lceil\Theta_r^4/K \rceil} - T_{\text{CSI}}\),
where \(\mathcal{C}_4^{'}\) is given by \eqref{cyc_MD4}. Hence, the cycle time for MAC Design 4 with retransmissions is given by 
\begin{equation}
\label{cyc_MD4_ret}
\mathcal{C}_4=\mathcal{C}_4^{'}+T_{\text{HIFS}}+T_{\text{Ret}}+\sum_{j=2}^{H} \Theta_{p,j}^4 \cdot T_{\text{Ret}}\vert_{\Theta_{r}^4=\Theta_{p,j}^4},
\end{equation}
where \(\Theta_{p,j}^4\) can be computed using \(P_{\text{fail,4}}^{o}\) in a similar way as described for previous MAC designs.

%
%
%
%

\begin{table}
\caption{Parameters for Performance Evaluation}
\label{table_param}
\vspace{-3ex}
\begin{center}
\begin{tabular}{ll}
\toprule
\bf{Parameter} & \bf{Value} \\
\midrule
Default no. of groups (\(M\)) & \(4\)\\
HIFS duration & \(40\) \(\mu\)s\\
Guard duration & \(10\) \(\mu\)s\\
Payload & \(64\) bytes\\
Minimum contention window & \(16\)\\
Carrier sensing threshold & \(-82\) dBm\\ 
Base rate (MAC Designs 1--3) & \(6\) Mbps \\
Enhanced rate (MAC Designs 1--3) & \(54\) Mbps \\
Header size (MAC Designs 1--3) & \(128\) bits (PHY), \(224\) bits (MAC) \\
Base rate (MAC Design 4) & \(16\) Mbps \\
PHY duration (MAC Design 4) & \(56\) \(\mu\)s \\
PRAW size & \(10\%\) of frame duration \\
Block ACK duration & \(31\) \(\mu\)s\\
CSI report duration & \(100\) \(\mu\)s \\
\textcolor{black}{Traffic pattern} & \textcolor{black}{Cyclic} \\

\hline
\end{tabular}
\end{center}
\end{table}

\section{\textsf{HAR\(^\text{2}\)D-Fi} -- Performance Evaluation} \label{sect_pe}
\textcolor{black}{We have evaluated the performance of \textsf{HAR\(^\text{2}\)D-Fi} through numerical (based on the analytic framework in Section \ref{sect_af}) and system-level simulation studies.  The simulation model considers \(N\) uniformly distributed STAs in the overlapping coverage of two access points. The channel model is characterized by large-scale path loss with \(\alpha=4\) and small-scale Rayleigh fading with \(\mu=1\). The customized simulator (in MATLAB) is based on IEEE 802.11 specifications and implements different MAC designs. The transmit power is set to \(23\) dBm and \(18\) dBm for the access points and the STAs, respectively. The SIFS, DIFS and PIFS duration is set to \(10\) \(\mu\)s, \(28\) \(\mu\)s and \(20\) \(\mu\)s, respectively. Other parameters are given in \tablename~\ref{table_param}.  We have conducted Monte Carlo simulations with a different user distribution and channel realization in each iteration. We have benchmarked the performance of \textsf{HAR\(^\text{2}\)D-Fi} against RT-WiFi and Wi-Red. Both of the baseline protocols have been implemented as per their specifications. Similar parameters have been used for all protocols. To ensure a fair comparison against time-slotted protocols like RT-WiFi and \textsf{HAR\(^\text{2}\)D-Fi}, we have eliminated any multi-user contention aspects for Wi-Red. }



\begin{figure}
\centering
\includegraphics[scale=0.2]{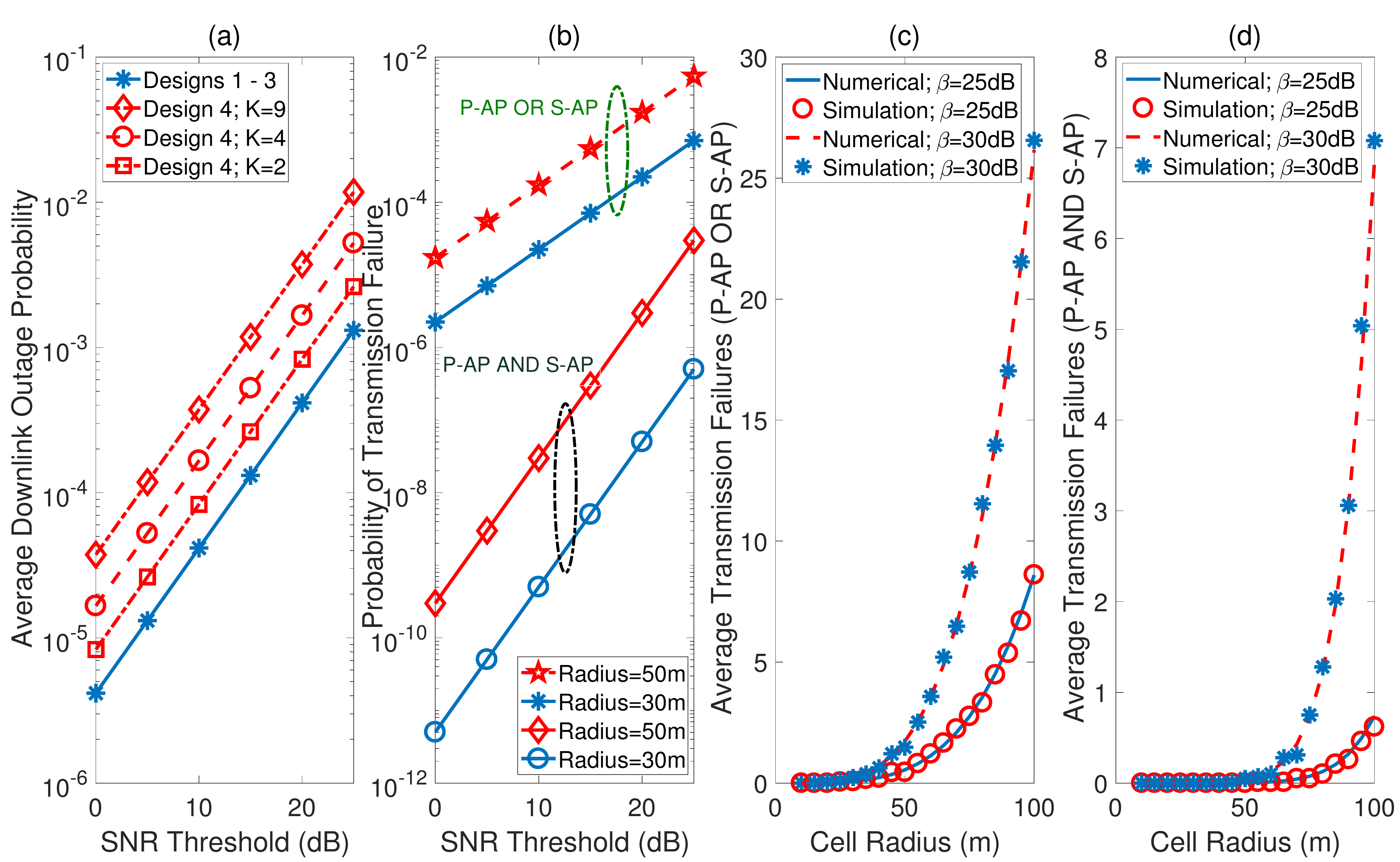}
\caption{Numerical results for (a) average downlink outage probability against SNR
threshold (\(\mathcal{R}_A=50\)m); (b) probability of transmission failure against SNR threshold for MAC Designs 1 -- 3. Results for average transmission failures (c) single access point (\(N=100\)); (d) both access points (\(N=100\)).}
\label{out_retrans}
\end{figure}

Fig. \ref{out_retrans}a compares the average downlink outage probability for different  designs, which  increases as the SNR requirements become more stringent.  Due to the multi-user nature of MAC Design 4, its outage probability is relatively higher. Fig. \ref{out_retrans}b shows the benefit of temporal redundancy as the probability of transmission failure with two access points is several orders of magnitude lower as compared to the single access point case. Figs. \ref{out_retrans}c and  \ref{out_retrans}d show the numerical and simulation results for average number of transmission failures, which increases as the cell radius or the SNR threshold increases due to higher outage probability. Note that temporal redundancy provides robustness to operate without any transmission failures over relatively larger cell radius as compared to the single access point case. The simulation results closely follow the numerical results and validate the analytic modeling.


\begin{figure}
\centering
\includegraphics[scale=0.2]{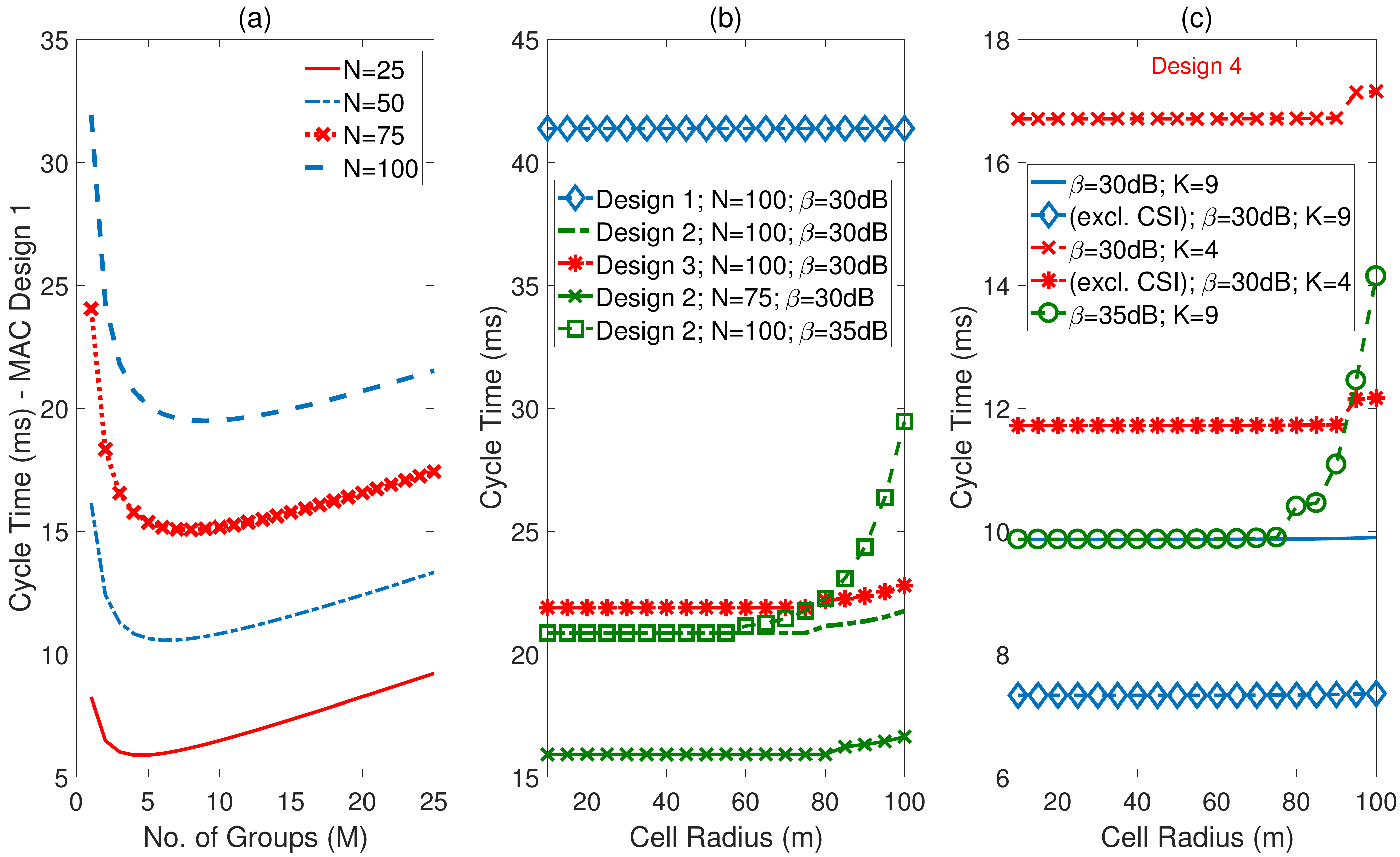}
\caption{Numerical results for (a) cycle time for MAC Design 1 against the number of groups (\(N_{\text{SF}}=1\)); (b) cycle time for MAC Designs 1 -- 3 against cell radius with \(H=3\); (c) Cycle time for MAC Design 4 against cell radius and \(H=3\). }
\vspace{-0.3cm}
\label{CT_numerical}
\end{figure}

Fig. \ref{CT_numerical} shows the numerical results for cycle time which capture the \textit{average} performance. Fig. \ref{CT_numerical}a evaluates the impact of number of groups (\(M\)) on cycle time of MAC Design 1. The cycle time initially decreases  as \(M\) increases. This is due to more parallel transmission opportunities. However, after a certain number of groups, the cycle time begins to increase due to higher overhead from frequent beacon transmissions and channel switching operation. Fig. \ref{CT_numerical}b shows the cycle time for MAC Designs 1 -- 3. With one superframe repetition, MAC Design 1 achieves a cycle time of \(41.3\) ms for \(100\) STAs. Other OFDM-based MAC designs implement PCF-based operation for handling retransmissions. The cycle time for MAC Designs 2 and 3 increases for larger cell radii due to higher number of retransmissions. For  \(\mathcal{R}_A=100\)m and \(\beta=30\) dB, MAC Designs 2 and 3 achieve a cycle time of \(21.6\) ms and \(22.8\) ms, respectively. With \(\beta=35\)dB, the cycle time of MAC design 2 increases to \(29.4\) ms. Fig. \ref{CT_numerical}c shows the cycle time of MAC Design 4. Due to multi-user transmissions, MAC Design 4 outperforms other MAC designs. With \(\beta=30\) dB and \(\mathcal{R}_A=100\)m, it achieves a cycle time of \(9.9\) ms and \(17.2\) ms, for \(9\) and \(4\) STAs per multi-user transmission, respectively. Note that this is the worst-case performance wherein CSI acquisition takes places every cycle. Without CSI acquisition every cycle, the cycle time reduces to \(7.4\) ms and \(12.1\) ms, respectively. With \(\beta=35\) dB and \(\mathcal{R}_A=100\)m, it achieves a cycle time of \(14.1\) ms with CSI acquisition. This is nearly \(52\%\) better than MAC Design 2 in a similar scenario.

\begin{figure}
\centering
\includegraphics[scale=0.2]{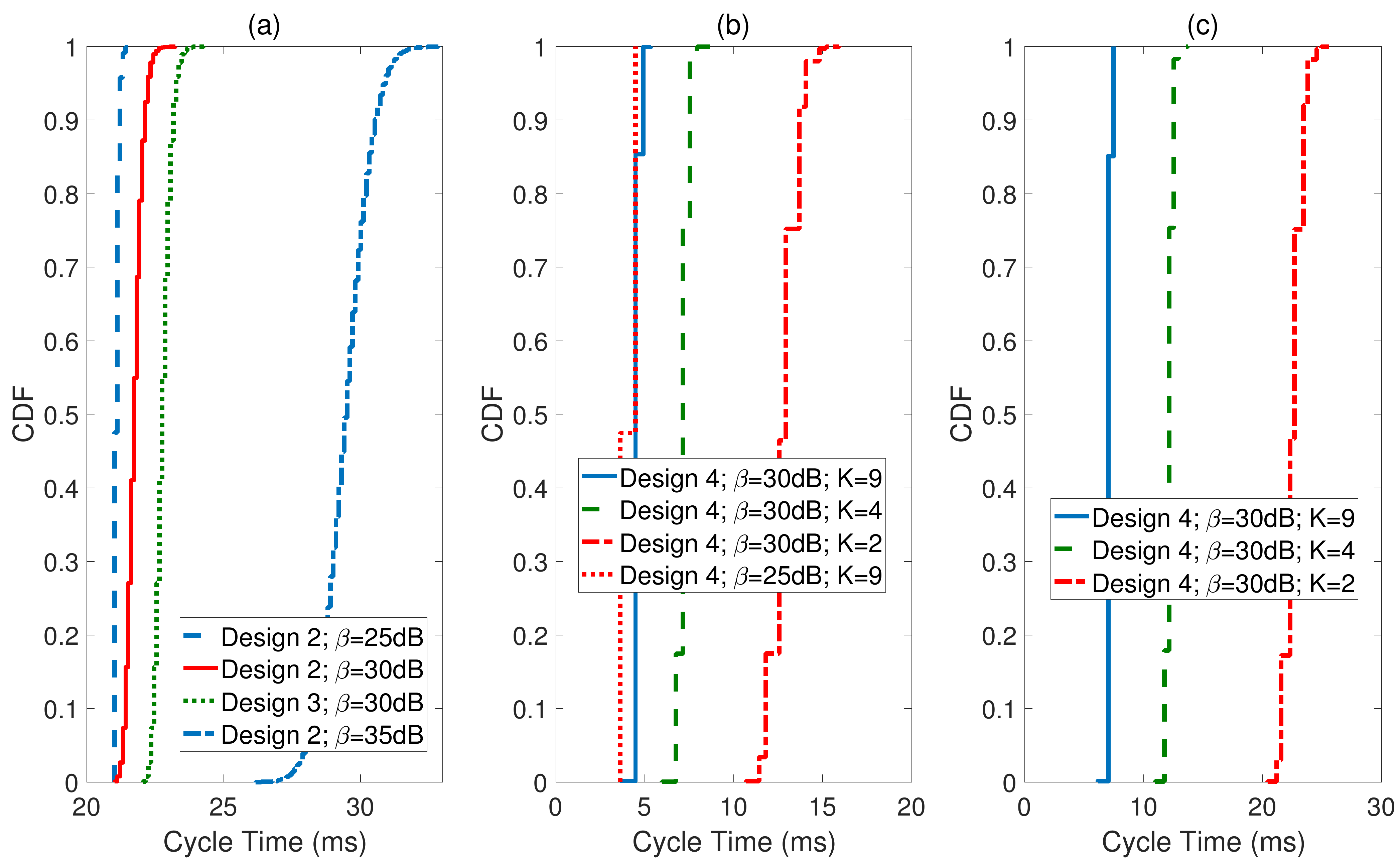}
\caption{Simulation results for the CDF (over \(100\) iterations) of cycle time with \(\mathcal{R}_A=100\)m; (a) MAC Designs 2 and 3; (b) MAC Design 4 (without CSI acquisition) with \(H=2\); (c) MAC Design 4 (with CSI acquisition) with \(H=2\); }
\label{CT_sim}
\end{figure}

%
%

Fig. \ref{CT_sim} shows the cumulative distribution function (CDF) of cycle time based on simulations. The simulation results show good similarity with the numerical results in Fig. \ref{CT_numerical}. For instance, with \(\beta=30\) dB and \(\mathcal{R}_A=100\)m, the CDF for MAC Design 2 shows a cycle time of \(21.1-21.9\) ms. The average cycle time for MAC Design 2 in a similar scenario is \(21.6\) ms. With \(\beta=35\) dB, the CDF shows a cycle time of \(27-32.5\) ms with a \(50\)th percentile of \(29.4\) ms which is the average cycle time. With \(\beta=30\) dB, \(\mathcal{R}_A=100\)m and \(9\) STAs per multi-user transmission, MAC Design 4 achieves a cycle time of \(3.9-5.4\) ms and \(6.1-7.9\) ms, without CSI acquisition and with CSI acquisition, respectively. With \(4\) STAs per multi-user transmission, a cycle time of \(6-8.8\) ms and \(10.9-13.7\) ms can be achieved in a similar scenario.

%

\begin{figure}
\centering
\includegraphics[scale=0.2]{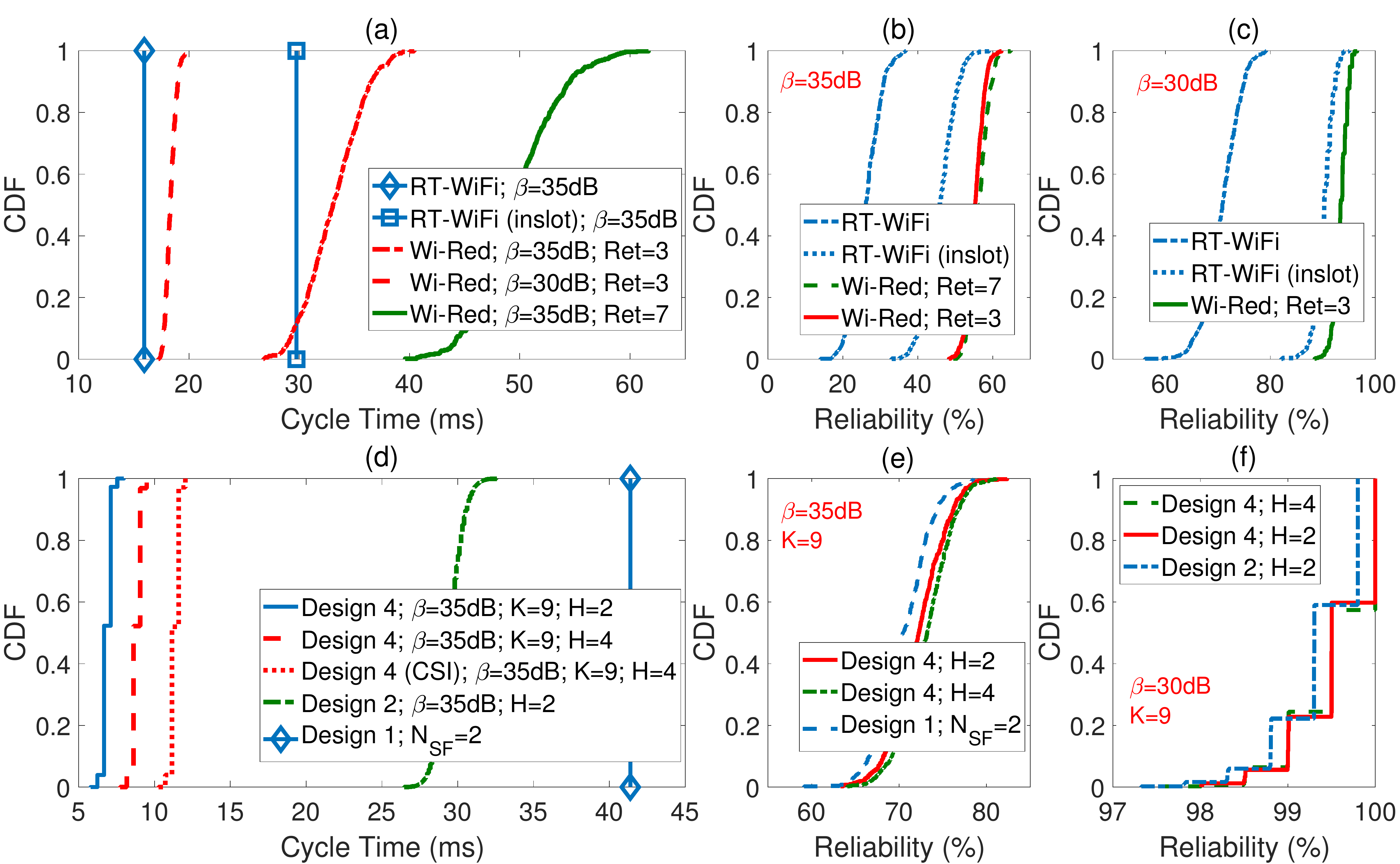}
\caption{Simulation results for baseline protocols and \textsf{HAR\(^\text{2}\)D-Fi} (CDF over \(100\) iterations, \(N=100\) and \(\mathcal{R}_A=100\)m); cycle time performance in (a) and (d); reliability performance (\(\beta=35\)dB) in (b) and (e);   reliability performance (\(\beta=30\)dB) in (c) and (f), respectively. }
\label{comp}
\end{figure}

Fig. \ref{comp} provides performance benchmarking of \textsf{HAR\(^\text{2}\)D-Fi} against RT-WiFi and Wi-Red in terms of cycle time and reliability (packet delivery). Figs. \ref{comp}a -- \ref{comp}c evaluate baseline protocols. With \(\beta=35\) dB, RT-WiFi achieves a cycle time of \(15.9\) ms which increases to \(29.7\) ms  with in-slot retransmissions. However, RT-WiFi provides insufficient reliability. With \(\beta=35\) dB, it provides a \(90\)th percentile reliability of \(30.8\%\), which increases to \(50.9\%\) with in-slot retransmission. With  \(\beta=30\) dB, it provides a \(90\)th percentile reliability of \(77.7\%\), which increases  to \(92.4\%\) with in-slot retransmission. With \(\beta=35\) dB, Wi-Red  achieves a  \(90\)th percentile cycle time of \(36.5\) ms and \(54.6\) ms  with a maximum of \(3\) and \(7\) MAC layer retransmissions, respectively. The corresponding reliability figures are \(58.1\%\) and \(60.5\%\), respectively. With \(\beta=30\) dB and \(3\) retransmissions, it achieves a \(90\)th percentile reliability of \(94.7\%\). 
Figs. \ref{comp}d -- \ref{comp}f show the performance of \textsf{HAR\(^\text{2}\)D-Fi}. With \(\beta=35\) dB and \(9\) STAs per multi-user transmission, MAC Design 4 achieves a \(90\)th percentile cycle time of \(7.1\) ms and \(9\) ms, with a maximum of \(2\) and \(4\) retransmission phases (\(H\)), respectively.  With CSI acquisition every cycle, the cycle time for the latter increases to \(11.5\) ms. Results show that MAC Design 4 outperforms  RT-WiFi and Wi-Red in terms of cycle time. Note the little variation in cycle time  which demonstrates deterministic performance of \textsf{HAR\(^\text{2}\)D-Fi}. Results also show that MAC Design 2 outperforms Wi-Red whereas RT-WiFi outperforms MAC Design 2 in terms of cycle time, in a similar setting. However, it is also important to look at the reliability performance. With \(\beta=35\) dB, MAC Design 4 achieves a \(90\)th percentile reliability of \(75.6\%\) and  \(77.2\%\) when \(H=2\) and \(H=4\), respectively. MAC Design 1 achieves a comparable reliability; however, at the expense of relatively higher cycle time performance. With \(\beta=30\) dB, MAC Design 4 as well as MAC Design 2 provide a reliability of \(97\%-100\%\) with a \(90\)th percentile of \(100\%\).  The results demonstrate that \textsf{HAR\(^\text{2}\)D-Fi} outperforms RT-WiFi (by up to \(35\%\)) and Wi-Red (by up \(26\%\))  in terms of reliability, particularly under challenging wireless environments. Both baseline protocols would incur significantly higher latency while providing a similar reliability as \textsf{HAR\(^\text{2}\)D-Fi}.

\section{Concluding Remarks} \label{sect_cr}
This paper introduced \textsf{HAR\(^\text{2}\)D-Fi} for control-centric industrial applications. \textsf{HAR\(^\text{2}\)D-Fi} overcomes the limitations of legacy Wi-Fi and provides deterministic and reliable connectivity through incorporation of hybrid channel access mechanisms and temporal redundancy techniques. Performance evaluation demonstrates that \textsf{HAR\(^\text{2}\)D-Fi} outperforms state-of-the-art Wi-Fi solutions in terms of latency and reliability. Results also demonstrate the superiority of OFDMA-based MAC design over other MAC designs. It achieves a cycle time of \(<10\) ms  for \(100\) devices with very high reliability of \(>99.75\%\) for up to \(100\)m cell radius.  Since the performance gain of \textsf{HAR\(^\text{2}\)D-Fi} comes without hardware complexity, it provides a cost-effective industrial networking solution. \textsf{HAR\(^\text{2}\)D-Fi} can be implemented on any off-the-shelf Wi-Fi chipset without requiring changes at the PHY layer.






\bibliographystyle{IEEEtran}

\bibliography{IEEEabrv,HFI_bib}

\begin{thebibliography}{10}
\providecommand{\url}[1]{#1}
\csname url@samestyle\endcsname
\providecommand{\newblock}{\relax}
\providecommand{\bibinfo}[2]{#2}
\providecommand{\BIBentrySTDinterwordspacing}{\spaceskip=0pt\relax}
\providecommand{\BIBentryALTinterwordstretchfactor}{4}
\providecommand{\BIBentryALTinterwordspacing}{\spaceskip=\fontdimen2\font plus
\BIBentryALTinterwordstretchfactor\fontdimen3\font minus
  \fontdimen4\font\relax}
\providecommand{\BIBforeignlanguage}[2]{{%
\expandafter\ifx\csname l@#1\endcsname\relax
\typeout{** WARNING: IEEEtran.bst: No hyphenation pattern has been}%
\typeout{** loaded for the language `#1'. Using the pattern for}%
\typeout{** the default language instead.}%
\else
\language=\csname l@#1\endcsname
\fi
#2}}
\providecommand{\BIBdecl}{\relax}
\BIBdecl

\bibitem{emer_ind_net}
J.~R. Moyne and D.~M. Tilbury, ``{The Emergence of Industrial Control Networks
  for Manufacturing Control, Diagnostics, and Safety Data},'' \emph{Proc.
  IEEE}, vol.~95, no.~1, pp. 29--47, January 2007.

\bibitem{ind_cont_net}
B.~Galloway and G.~Hancke, ``{Introduction to Industrial Control Networks},''
  \emph{IEEE Commun. Surveys Tuts.}, vol.~15, no.~2, pp. 860--880, Second
  Quarter 2013.

\bibitem{ind4}
R.~Drath and A.~Horch, ``{Industrie 4.0: Hit or Hype?}'' \emph{IEEE Ind.
  Electron. Mag.}, vol.~8, no.~2, pp. 56--48, June 2014.

\bibitem{TI_PIEEE}
A.~Aijaz and M.~Sooriyabandara, ``{The Tactile Internet for Industries: A
  Review},'' \emph{Proc. IEEE}, vol. 107, no.~2, pp. 414--435, Feb. 2019.

\bibitem{fieldbus}
J.-P. Thomesse, ``{Fieldbus Technology in Industrial Automation},'' \emph{Proc.
  IEEE}, vol.~93, no.~6, pp. 1073--1101, June 2005.

\bibitem{ind_ethernet}
J.-D. Decotignie, ``{Ethernet-Based Real-Time and Industrial Communications},''
  \emph{Proc. IEEE}, vol.~93, no.~6, pp. 1102--1117, June 2005.

\bibitem{enclose}
A.~{Aijaz}, ``{\textsf{ENCLOSE}: An Enhanced Wireless Interface for
  Communication in Factory Automation Networks},'' \emph{IEEE Trans. Ind.
  Informat.}, vol.~14, no.~12, pp. 5346--5358, Dec 2018.

\bibitem{WISA}
G.~{Scheible}, D.~{Dzung}, J.~{Endresen}, and J.~E. {Frey}, ``{Unplugged but
  Connected [Design and Implementation of a Truly Wireless Real-Time
  Sensor/Actuator Interface]},'' \emph{IEEE Ind. Electron. Mag.}, vol.~1,
  no.~2, pp. 25--34, Summer 2007.

\bibitem{gallop}
A.~Aijaz, A.~Stanoev, and U.~Raza, ``{GALLOP: Toward High-Performance
  Connectivity for Closing Control Loops over Multi-Hop Wireless Networks},''
  in \emph{ACM International Conference on Real-Time Networks and Systems
  (RTNS)}, 2019, p. 176–186.

\bibitem{f_ind_comm}
M.~Wollschlaeger, T.~Sauter, and J.~Jasperneite, ``{The Future of Industrial
  Communication: Automation Networks in the Era of the Internet of Things and
  Industry 4.0},'' \emph{IEEE Ind. Electron. Mag.}, vol.~11, no.~1, pp. 17--27,
  March 2017.

\bibitem{11ax}
\BIBentryALTinterwordspacing
``{IEEE 802.11ax High Efficiency WLAN (HEW)},'' accessed: 2019-11-20. [Online].
  Available: \url{http://www.ieee802.org/11/Reports/tgax_update.htm}
\BIBentrySTDinterwordspacing

\bibitem{IEC_PRP}
\BIBentryALTinterwordspacing
IEC, ``{Industrial Communications Networks -- High Availability Automation
  Networks -- Part 3: Parallel Redundancy Protocol (PRP) and High-Availability
  Seamless Redundancy (HSR)},'' {International Electrotechnical Commission
  (IEC)}, Standard (2nd Edition) {62439-3}, 2012. [Online]. Available:
  \url{https://webstore.iec.ch/publication/24447}
\BIBentrySTDinterwordspacing

\bibitem{Wi-Red}
G.~Cena, S.~Scanzio, and A.~Valenzano, ``{Seamless Link-Level Redundancy to
  Improve Reliability of Industrial Wi-Fi Networks},'' \emph{IEEE Trans. Ind.
  Informat.}, vol.~12, no.~2, pp. 608--620, April 2016.

\bibitem{rt_wifi}
Y.-H. Wei \emph{et~al.}, ``{RT-WiFi: Real-Time High-Speed Communication
  Protocol for Wireless Cyber-Physical Control Applications},'' in \emph{IEEE
  Real-Time Systems Symposium (RTSS)}, Dec. 2013, pp. 140--149.

\bibitem{soft_tdmac}
P.~Djukic and P.~Mohapatra, ``{Soft-TDMAC: A Software TDMA-based MAC over
  Commodity 802.11 hardware},'' in \emph{IEEE International Conference on
  Computer Communications (INFOCOM)}, April 2009, pp. 1836--1884.

\bibitem{det_wifi}
Y.~Cheng, D.~Yang, and H.~Zhou, ``{Det-WiFi: A Multihop TDMA MAC Implementation
  for Industrial Deterministic Applications Based on Commodity 802.11
  Hardware},'' \emph{Wireless Commun. and Mobile Comput.}, pp. 1--10, April
  2017.

\bibitem{S_IWLAN}
``{Siemens SCALANCE W},''
  \url{http://w3.siemens.com/mcms/industrial-communication/en/industrial-wireless-communication/iwlan-industrial-wireless-lan/pages/iwlan.aspx},
  accessed: 2019-05-10.

\bibitem{802_11e}
``{IEEE Standard for Information technology-- Local and metropolitan area
  networks--Specific requirements--Part 11: Wireless LAN Medium Access Control
  (MAC) and Physical Layer (PHY) Specifications - Amendment 8: Medium Access
  Control (MAC) Quality of Service Enhancements},'' \emph{IEEE Std
  802.11e-2005}, Nov. 2005.

\bibitem{11ax_RM}
M.~Karaca \emph{et~al.}, ``{Resource Management for OFDMA based Next Generation
  802.11 WLANs},'' in \emph{IFIP Wireless and Mobile Networking Conference
  (WMNC)}, July 2016.

\bibitem{11ax_S}
D.~Bankov \emph{et~al.}, ``{IEEE 802.11ax Uplink Scheduler to Minimize Delay: A
  Classic Problem with New Constraints},'' in \emph{IEEE International
  Symposium on Personal, Indoor and Mobile Radio Communications (PIMRC)}, Oct.
  2017.

\bibitem{w_low_fa}
F.~D. Pellegrini, D.~Miorandi, S.~Vitturi, and A.~Zanella, ``{On the Use of
  Wireless Networks at Low Level of Factory Automation Systems},'' \emph{IEEE
  Trans. Ind. Informat.}, vol.~2, no.~2, pp. 129--143, May 2006.

\bibitem{802_11_ah}
``{IEEE Standard for Information technology--Telecommunications and information
  exchange between systems - Local and metropolitan area networks--Specific
  requirements - Part 11: Wireless LAN Medium Access Control (MAC) and Physical
  Layer (PHY) Specifications Amendment 2: Sub 1 GHz License Exempt
  Operation},'' \emph{IEEE Std 802.11-2007 (Revision of IEEE Std 802.11-1999)},
  May 2017.

\bibitem{dist_reg}
Z.~Khalid and S.~Durrani, ``{Distance Distributions in Regular Polygons},''
  \emph{IEEE Trans. Veh. Technol.}, vol.~62, no.~5, pp. 2363--2368, Jun 2013.

\end{thebibliography}
%

\begin{IEEEbiographynophoto}{Adnan Aijaz}
(M'14 -- SM'18) is a Principal Research Engineer with the Bristol Research and Innovation Laboratory of Toshiba where he is actively engaged in industrial wireless research and technology development activities. He has contributed to various  international research projects and standardization activities. He holds a Ph.D in telecommunications engineering from King’s College London, United Kingdom. He has a number of publications and patents in the area of wireless communications and networking.
\end{IEEEbiographynophoto}

\end{document}